\documentclass[prl,twocolumn,aps,showpacs,superscriptaddress]{revtex4}
\usepackage{graphics}
\usepackage{epsfig}
\usepackage{amsmath}
\usepackage{amssymb}
\usepackage{bbm}
\usepackage{ulem}

\begin{document}
\newcommand{\beq}{\begin{equation}}
\newcommand{\eeq}{\end{equation}}
\newcommand{\legende}[2]{\caption[#1]{\label{#1}{#2}}}
\newcommand{\ket}[1]{\left|#1\right>}
\newcommand{\bra}[1]{\left<#1\right|}
\newcommand{\braket}[2]{\left<#1\middle| #2 \right>}

\title{Soft Sphere Packings at Finite Pressure but Unstable to
Shear}

\author{Simon Dagois-Bohy}
\affiliation{Kamerling Onnes Lab, Universiteit Leiden, Postbus
9504, 2300 RA Leiden, The
Netherlands}\affiliation{Instituut-Lorentz, Universiteit Leiden,
Postbus 9506, 2300 RA Leiden, The Netherlands}
\author{Brian P.~Tighe}
\affiliation{Instituut-Lorentz, Universiteit Leiden, Postbus 9506,
2300 RA Leiden, The Netherlands} \affiliation{Delft University of
Technology, Process \& Energy Laboratory, Leeghwaterstraat 44 2628
CA Delft, The Netherlands }
\author{Johannes Simon}
\affiliation{Kamerling Onnes Lab, Universiteit Leiden, Postbus
9504, 2300 RA Leiden, The Netherlands}
\author{Silke Henkes}
\affiliation{Instituut-Lorentz, Universiteit Leiden, Postbus 9506,
2300 RA Leiden, The Netherlands}\affiliation{Physics Department,
Syracuse University, Syracuse, NY 13244, USA}
\author{Martin van Hecke}
\affiliation{Kamerling Onnes Lab, Universiteit Leiden, Postbus
9504, 2300 RA Leiden, The Netherlands}

\date{\today}

\begin{abstract}
When are athermal soft sphere packings jammed? Any experimentally
relevant definition must at the very least require a jammed
packing to resist shear. We demonstrate that widely used
(numerical) protocols in which particles are compressed together,
can and do produce packings which are unstable to shear --- and
that the probability of generating such packings reaches one near
jamming. We introduce a new protocol that, by allowing the system
to explore different box shapes as it equilibrates, generates
truly jammed packings with strictly positive shear moduli $G$. For
these packings, the scaling of the average of $G$ is consistent
with earlier results, while the probability distribution $P(G)$
exhibits novel and rich scalings.
\end{abstract}

\maketitle

Foams, emulsions, colloidal suspensions, granular media and other
particulate media undergo a jamming transition when their
constituent particles are packed densely enough
\cite{majmudar07,lechenault08,jenkins08,clusel09,katgert10b,cheng10,metayer11}.
This transition has been extensively studied in packings of
deformable, athermal, frictionless spheres interacting through
purely repulsive contact forces
\cite{bolton90,durian95,ohern03,vanhecke10,liu10a}. The limit where
the particles just touch then plays the role of an unusual
critical point, as a host of quantities, such as shear modulus,
time and length scales, and contact number exhibit power law
scaling with the distance to this critical point
\cite{bolton90,durian95,ohern03,silbert05,zhang05,ellenbroek06,vanhecke10,liu10a,tighe11b,tighe11}.

Numerically created particle packings play a central role in many
fields of physics, in particular jamming. In all numerical jamming
studies we are aware of, packings are created by compressing a
collection of particles, either by inflating the particles or
shrinking the simulation box
\cite{bolton90,durian95,ohern03,silbert05,zhang05,ellenbroek06,vanhecke10,liu10a,tighe11b,tighe11}.
It is then widely believed and tacitly assumed that, when
compressed, the system simultaneously develops a finite pressure,
a finite yield threshold \cite{durian95,ohern03} and a positive
shear modulus $G$
\cite{bolton90,durian95,ohern03,zhang05,vanhecke10,liu10a}. Here
we demonstrate that, to the contrary, algorithms that work solely
by compression tend to produce packings that are unstable to
shear, and hence have negative shear moduli.  Nevertheless, such
`improperly jammed' packings possess a positive pressure $P$ and a
positive bulk modulus, and are in mechanical equilibrium --- see
Fig.~1a.

In this Letter, we probe and explain this anomaly. The root
problem is that compression only (CO) algorithms ignore the global
shear degrees of freedom. We find that this results in a fraction
of improperly jammed CO packings which reaches {\it one} at the
critical point. Hence, compression alone does not lead to jammed
packings, and previous results on jamming have considered packings
that, instead of being jammed, have been linearly unstable to
shear --- in particular near jamming.

Furthermore, we remedy this anomaly by introducing a shear
stabilized (SS) packing algorithm that produces truly jammed
packings with positive definite shear moduli \cite{footnote1}, and
probe the probability distribution of $G$, uncovering novel
scaling with distance to jamming and system size.


\begin{figure}[tb]
\includegraphics[clip,width=0.9\linewidth]{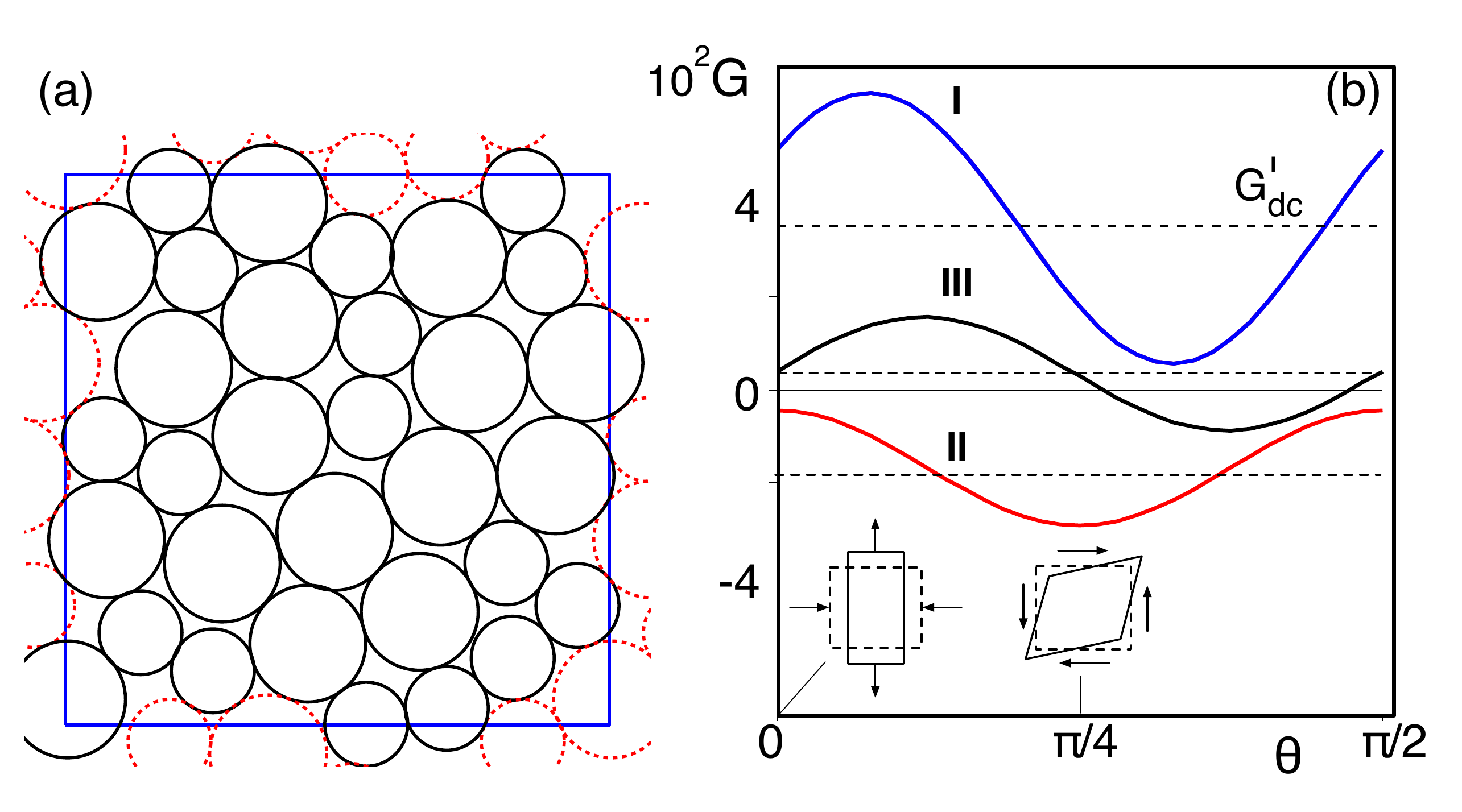}
\legende{Three_packs}{(Color online) (a) Example of a
well-equilibrated CO packing of $N=32$ particles which is unstable
to shear (pressure $P=10^{-2}$, bulk modulus  $K\approx  0.385$,
contact number $z\approx 4.26$). (b) Illustration of the
sinusoidal angular dependence of $G$ on the principle direction of
shear, $\theta$, for three different packings at the same $N$ and
$P$ --- curve III corresponds to the packing shown in (a), and
dashed lines indicate $G_{\rm DC}$, the angular average of $G$. }
\end{figure}

{\it Shear moduli in CO Packings ---} We have generated 2D
packings of $N$ soft harmonic bidisperse disks (with unit spring
constant \cite{vanhecke10}) by a standard CO packing generating
algorithm, for pressures $P$ ranging from $10^{-6}$ to $10^{-1}$
and $16\leq N \leq 1024$. Prior studies of the shear modulus have
focused on ensemble averages at fixed distance to the jamming
point ($P$), typically for large $N$, and without reference to the
angular dependence of $G$.

As illustrated in Fig.~1b, fluctuations and anisotropy are key:
$G$ varies sinusoidally with $\theta$, and its angular average,
$G_{\rm DC}$, varies substantially with realization. We
distinguish three types of packings. (I) Truly jammed packings for
which $G(\theta)>0$. (II) Improperly jammed packings for which
$G(\theta)<0$ (III) Improperly jammed packings for which
$G(\theta)$ becomes negative over an interval in $\theta$.
We stress that all these packings are in a mechanical equilibrium and have a
positive bulk modulus.

It has been customary to measure G along a fixed direction
\cite{ohern02,ohern03,ellenbroek06,maloney06,somfai07,wyart08,zaccone11},
and the limited unstable range of type III packings, combined with
the rare occurrence of type II packings, may explain why these
instabilities have escaped attention to date. Since simulations
often produce some ``problematic'' packings (for example due to
issues with convergence), packings of types II and III have likely
been treated as ``bad apples'' and thrown out of the ensemble
\cite{privcoms,schreck10}.

\begin{figure}[tb]
\includegraphics[width=.7\linewidth]{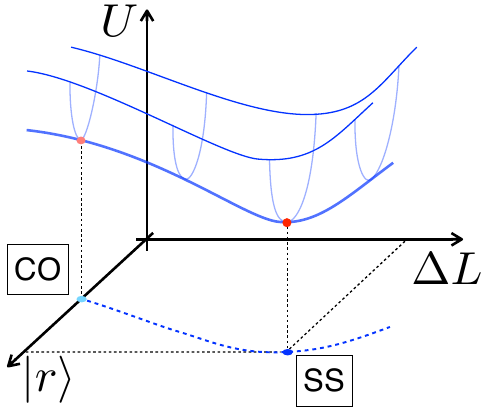}
\legende{Gpos}{(Color online) Energy landscape where $|r\rangle$
denotes the particle degrees of freedom, and $\Delta L$ the
box-shape. CO packings sit at a minimum of $U$ with respect to
$|r\rangle$; SS packings sit at a minimum with respect to both
$|r\rangle$ and $\Delta L$.}
\end{figure}

{\it Boundaries and Shear Stabilization ---} Improperly jammed packings are not
caused by numerical artifacts but stem from the essence of
compression only (CO) algorithms. Consider the potential energy
landscape as a function of the particle positions, $|r\rangle$,
and shear deformations of the box, $|\Delta L \rangle$ (Fig.~2).
CO algorithms fix the unit cell and generate packings at a minimum
of $U$ with respect to $|r\rangle$. Residual shear stresses and
shear moduli correspond to the first and second derivatives,
respectively, of $U$ along a strain direction $\Delta L$
--- without permitting the strain degrees of freedom to equilibrate,
both the residual stress and shear modulus are uncontrolled.

To create packings that are
guaranteed to be stable against shear in all directions, we
include shear deformations of the box and search for local energy
minima of $U$ (Fig.~2) \cite{notemin}. We combine standard
conjugate gradient techniques \cite{schreck10} with the FIRE
algorithm \cite{fire}, which improves the speed by an order of
magnitude, and also precisely control the pressure of the
resulting packings. Since the energy is at a minimum with respect to
the shear degrees of freedom, these packings have strictly
positive values of $G$ and exhibit zero residual shear stress
\cite{notemin}, unlike CO states. However, as a result of
equilibrating the strain degrees of freedom, the unit cell is no
longer square. For example, starting from a CO packing (minimum of
$U$ with respect to $|r\rangle$), the box is deformed to find a
minimum in the extended space spanned by $|r\rangle$ and the
strain coordinates (Fig.~2). Such deformations are small for large
systems \cite{footnote3b}.

A formal way of capturing the role of the boundaries is in terms
of the stiffness matrices $\hat{K}^0$ and $\hat{K}$, where $\hat
K^0$ is the usual Hessian, while the ``extended Hessian'' $\hat
K$, introduced in a different context in Ref.~\cite{tighe11},
includes the dependence on the shear degrees of freedom --- for
details see the supplementary material. It can then be shown that
$G(\theta)$ is positive definite for all $\theta$  if all
eigenvalues of $\hat K$ are {\it positive} (excluding the trivial
zero energy translational modes). Defining $\lambda_{\rm{min}}$ as
the minimal eigenvalue of $\hat K$, the sufficient condition for a
packing to be stable against shear is $\lambda_{\rm{min}}>0$. In
contrast, a positive spectrum for the usual Hessian $\hat K^0$
only guarantees stability in a box with fixed boundaries, but does
not guarantee stability to all possible shear deformations (Fig.~1
and 2), contrary to the claim in Ref.~\cite{ohern04}.

\begin{figure}[tb]
\includegraphics[width=1\linewidth]{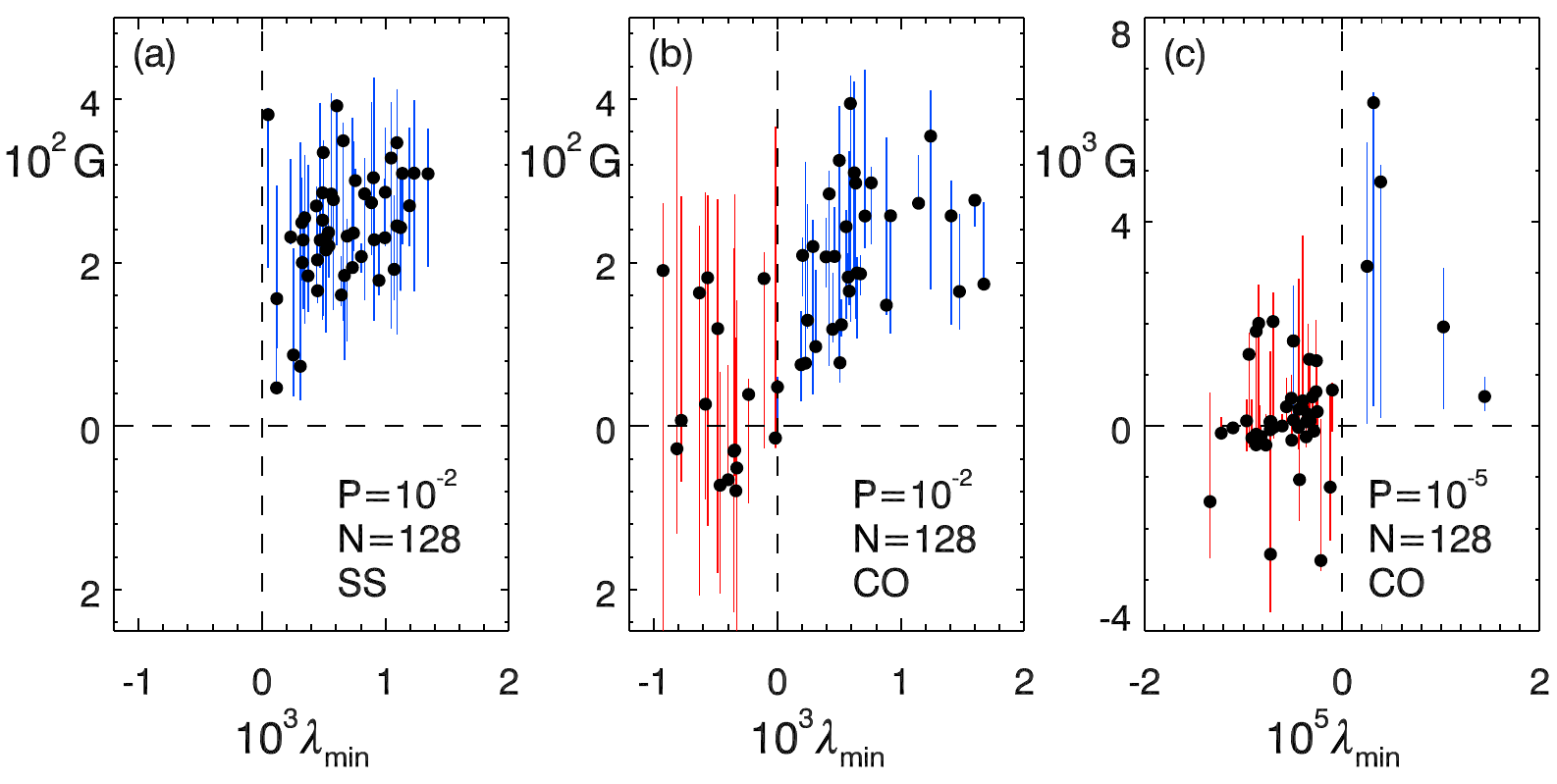}
\legende{quad}{(Color online) Scatter plots of $\lambda_{\rm min}$
vs $G$ for 50 packings of $N=128$ and $P$ as indicated. Dots
correspond to $G(\theta=0)$, and blue (red) lines indicate the
range of $G(\theta)$ when the minimum of $G(\theta)$ is positive
(negative). The right bottom quadrant is empty: when $\lambda_{\rm
min}>0$, $G$ is positive definite. (a) SS packings. (b) CO
packings at $P=10^{-2}$. (c) CO packings at $P=10^{-5}$ --- close
to jamming, the fraction of improperly jammed CO packings grows
dramatically.}
\end{figure}

Scatter plots of shear modulus and $\lambda_{\rm{min}}$ for CO and
SS ensembles shown in Fig.~3 confirm our picture: {\it(i)} All SS
packings have positive $\lambda_{\rm{min}}$ and $G$. {\it (ii)} CO
packings can have negative $\lambda_{\rm{min}}$. Although many of
these $\lambda_{\rm{min}}<0$ packings are stable when sheared
along a fixed direction (dots correspond to $\theta=0$), they
almost always have negative $G$ when sheared along other
directions.

{\it Fraction of improperly jammed CO packings ---} What fraction
of CO packings is unstable to shear? What governs the scaling of
this fraction? Fig.~4 shows that the probability that CO packings
have shear directions along which $G$ is negative, $P_{G<0}$ ,
reaches one near jamming, and that larger packings need lower
pressures for these instabilities to become dominant. It is
natural to expect that $P_{G<0}$ would collapse when plotted as a
function of $L /l^*$, where $l^*$ is a characteristic length-scale
which diverges as $1/\Delta z$ near jamming, and where $\Delta z$
is the difference between the contact number $z$ and its value at
the jamming point
\cite{tkachenko99,wyart05,ellenbroek06,vanhecke10,liu10a,Pnote}.
Surprisingly, Fig.~4 shows that the number of excess contacts
$\sim N \Delta z$, not the characteristic length scale $l^*$,
governs the fraction of improperly jammed packings --- note that
we have included a finite size correction to $\Delta z$ (see
below).

We conclude that the standard view of the jamming transition, in
which rigidity is attained by simply compressing particles
together \cite{ohern03,liu10a,vanhecke10}, needs a revision:
when the pressure is lowered in {\it finite} CO packings, more and
more packings will become unstable to shear, leading to a blurring
of the (un)jamming transition. We stress that many excess contacts
are needed to avoid improperly jammed CO packings: for example,
one needs of the order of a hundred excess contacts for
$P_{G<0}<0.1$.

\begin{figure}[tb]
\includegraphics[clip,width=0.9\linewidth]{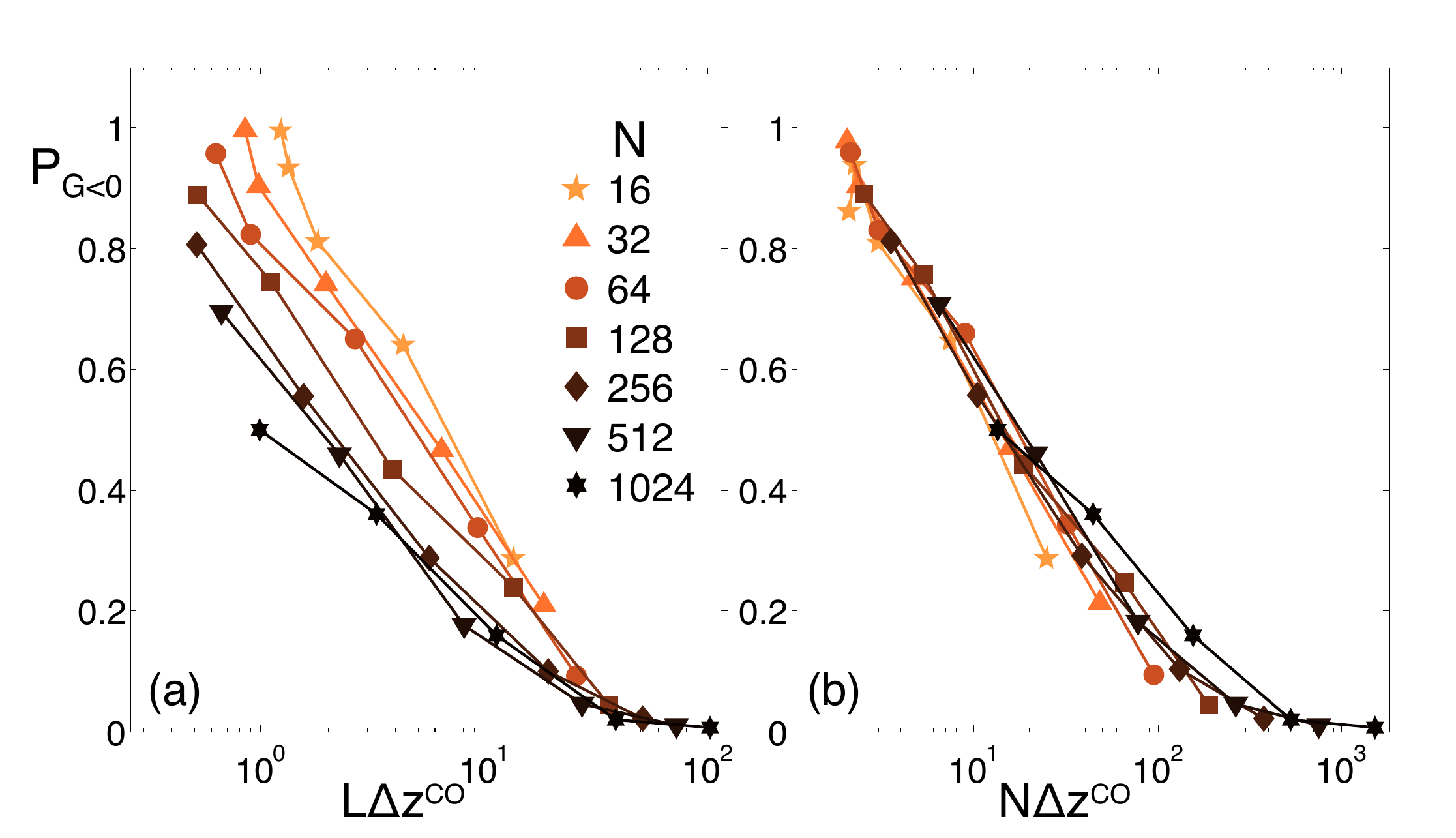}
\legende{jonagold}{(Color online) The fraction of CO packings
unstable to shear collapses when plotted as function of the excess
number of contacts, $N \Delta z^{\rm CO}$, where $\Delta  z^{\rm
CO}:= z-z_{\rm iso}^{\rm CO}=z-4+4/N$.}
\end{figure}

{\it Scaling of Contact Number and $G$ ---} Do the same scaling
laws for, e.g., $z$ or $G$ \cite{vanhecke10,liu10a}, govern both
CO and SS packings? To answer this question, we have performed a
finite size scaling analysis of both SS and CO packings: both the
distance to jamming and the system size play a crucial role
\cite{inprep}.

We first consider the contact number $z$
\cite{alexander,durian95,ohern03,liu10a,vanhecke10}. A packing is
called isostatic when the number of constraints, $C$, equals
$N_{\rm dof} - N_0$, the number of degrees of freedom $N_{\rm
dof}$ minus the number of rigid body modes $N_0$. There is one
constraint for each of the $N_c \equiv N z/2$ force bearing
contacts \cite{note_rattler}. In two dimensions, $N_0\!=\!2$,
corresponding to two rigid body translations (rotation is
incompatible with periodic boundary conditions). Hence:
\begin{equation}
z_{\rm iso} \ge \frac{2}{N}(N_{\rm dof}  - N_0) \,.
\label{eqn:isostatic}
\end{equation}
For CO states in two dimensions, $N_{\rm dof} = 2N$ (the particle
displacements), so that $z^{\rm CO}_{\rm iso} = 4 - 4/N$. For SS
states the particle displacements are augmented by two shear
degrees of freedom, leading to $z_{\rm iso}^{\rm SS} = 4$.

Is the isostatic bound reached at unjamming? We have found that
both CO and SS packings
%
have one contact in excess of their respective isostatic values
when approaching the jamming point (see Suppl.~Mat.). Goodrich
{\it et al.}~have argued that this extra contact reflects the
requirement that jammed states have positive bulk modulus, which
puts an additional constraint on the box size \cite{CSL}.

We now turn our attention to the scaling of $G$, and first
investigate the scaling of the angle-averaged shear modulus
$\langle G \rangle$ in ensembles of finite sized  CO and SS
packings. In Fig.~5a we show that in the CO ensemble, $\langle G
\rangle$ is proportional to $z - z_{\rm iso}^{\rm CO}$, consistent
with prior results \cite{ohern03,ellenbroek06,zaccone11,tighe11,CSL}.
In Fig.~5b we show that in the SS
ensemble, the average shear modulus is proportional to $z -
(z_{\rm iso}^{\rm SS} - 8/N)$.  So, although the SS shear modulus
is also linear in $z$, its vanishing point extrapolates to a state
with four contacts less than the isostatic state. We note that in
both ensembles $\langle G \rangle$ is of order $1/N$ in the zero
pressure limit.

\begin{figure}[tb]
\includegraphics[clip,width=0.9\linewidth]{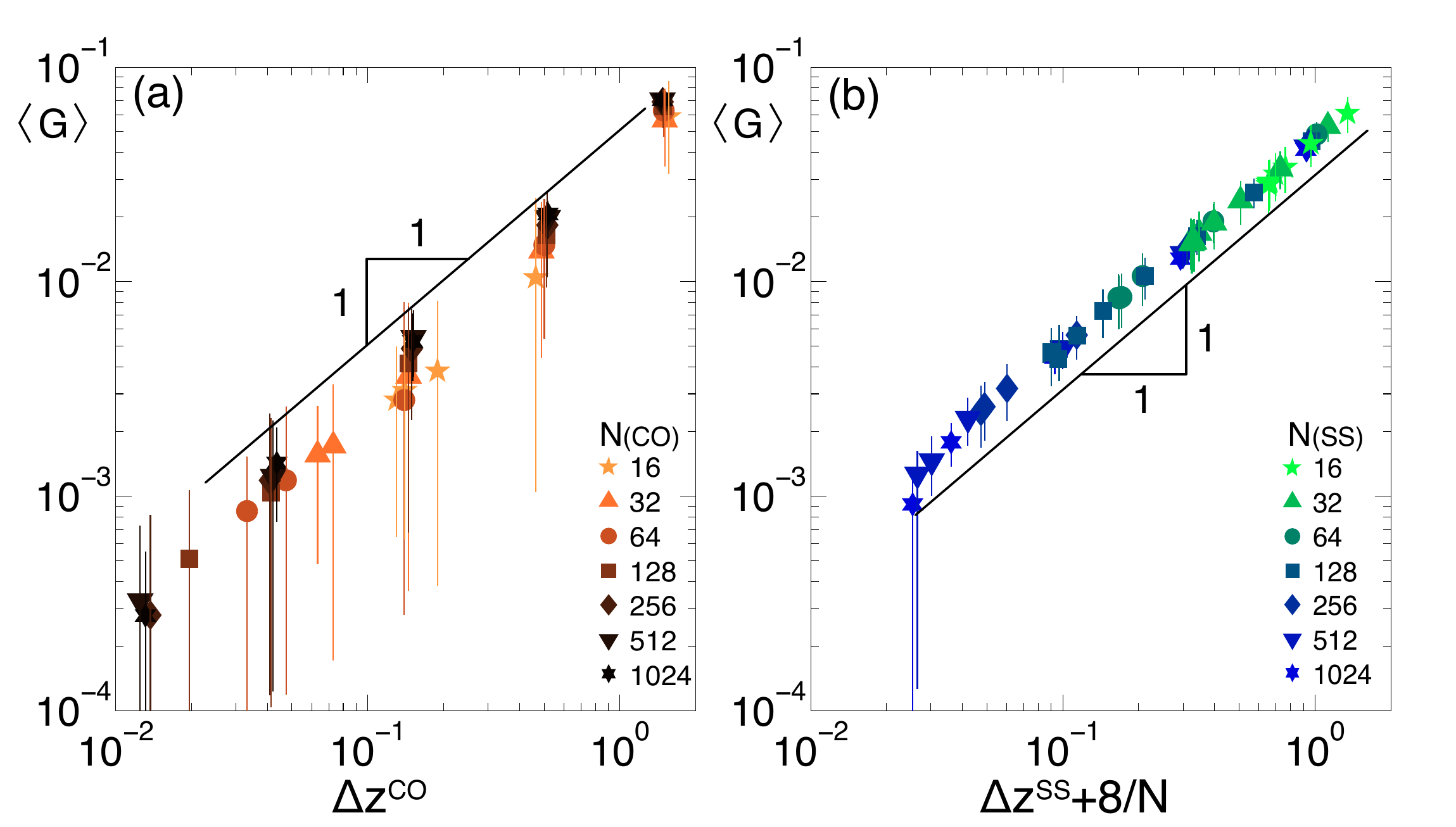}
\centering \legende{G}{(Color online) (a) Linear scaling of
$\langle G \rangle$ with $\Delta z^{\rm CO}$ for CO packings. The
errorbars indicate the RMS fluctuations in $G$. (b) Linear scaling
of $\langle G \rangle$ with $\Delta z^{\rm SS} \!+\! 8/N$ for SS
packings --- where  $\Delta z^{\rm SS}:= z-z_{\rm iso}^{\rm
SS}=z-4$. }
\end{figure}

The amount of scatter in $\langle G \rangle$ observed in our new
CO packings is surprisingly large. We note that previous work did
not consider the value of $G$ over all angles and discarded
negative values of $G$, which leads to a smaller scatter
\cite{privcoms,schreck10}. Recent work by Goodrich{\it et al.}
shows that this scatter can be further suppressed by using
exceptionally accurate equilibration and larger ensembles
\cite{CSL}. Nevertheless, the observation that SS data exhibits
far lower scatter than CO data, while both packings were obtained
with the same numerical accuracy, suggests that remnants of the
unstable modes present in the CO ensemble hinder accurate
equilibration.

With few exceptions
\cite{ohern03,henkes07,ellenbroek06,tighe08b,ellenbroek09,heussinger10,tighe10c,tighe11b},
studies of jamming have focused on ensemble averages. Here we
consider the probability distribution $P(G)$ for both ensembles,
sampling both $\theta$ and realizations. Fig.~6a illustrates that
for CO packings, $P(G)$ often peaks at {\it negative} $G$, and can
possess an extended tail towards negative $G$. In contrast, for SS
packings,  $G$ is strictly positive, and the peak of $P(G)$ is
always at finite $G$.

For SS packings, the distributions $P(G)$ are well-behaved;
however, there is no single parameter scaling. For brevity of
notation, we define $\tilde{z} \equiv 4-8/N$, so that $\langle G
\rangle \sim z - \tilde{z} \equiv \Delta \tilde{z}$. Our data
shows that the variance $\sigma_G^2$ scales roughly linear with
$\Delta \tilde{z}/L$ (Fig.~6b). The scalings of the average and
variance of $G$ suggest that distributions of $P(G/\langle G
\rangle)$ that have equal values of $L \Delta \tilde{z}$ might
collapse. Fig.~6c shows that grouping $P(G/\langle G \rangle)$ by
$L \Delta \tilde{z}$ captures the main trends: for large $L \Delta
\tilde{z}$, the distribution $P(G/\langle G \rangle)$ is clearly
peaked away from zero, but for lower values of $L \Delta
\tilde{z}$ becomes more skewed and wider. We note that the scaling
of $P(G<0)$ for CO packings suggest that finite size scaling
corrections for $P(G)$ differ between CO and SS packings, and an
important question for the future is to probe these differences
\cite{carlpriv}.

\begin{figure}[tb]
\centering
\includegraphics[width=1.\linewidth]{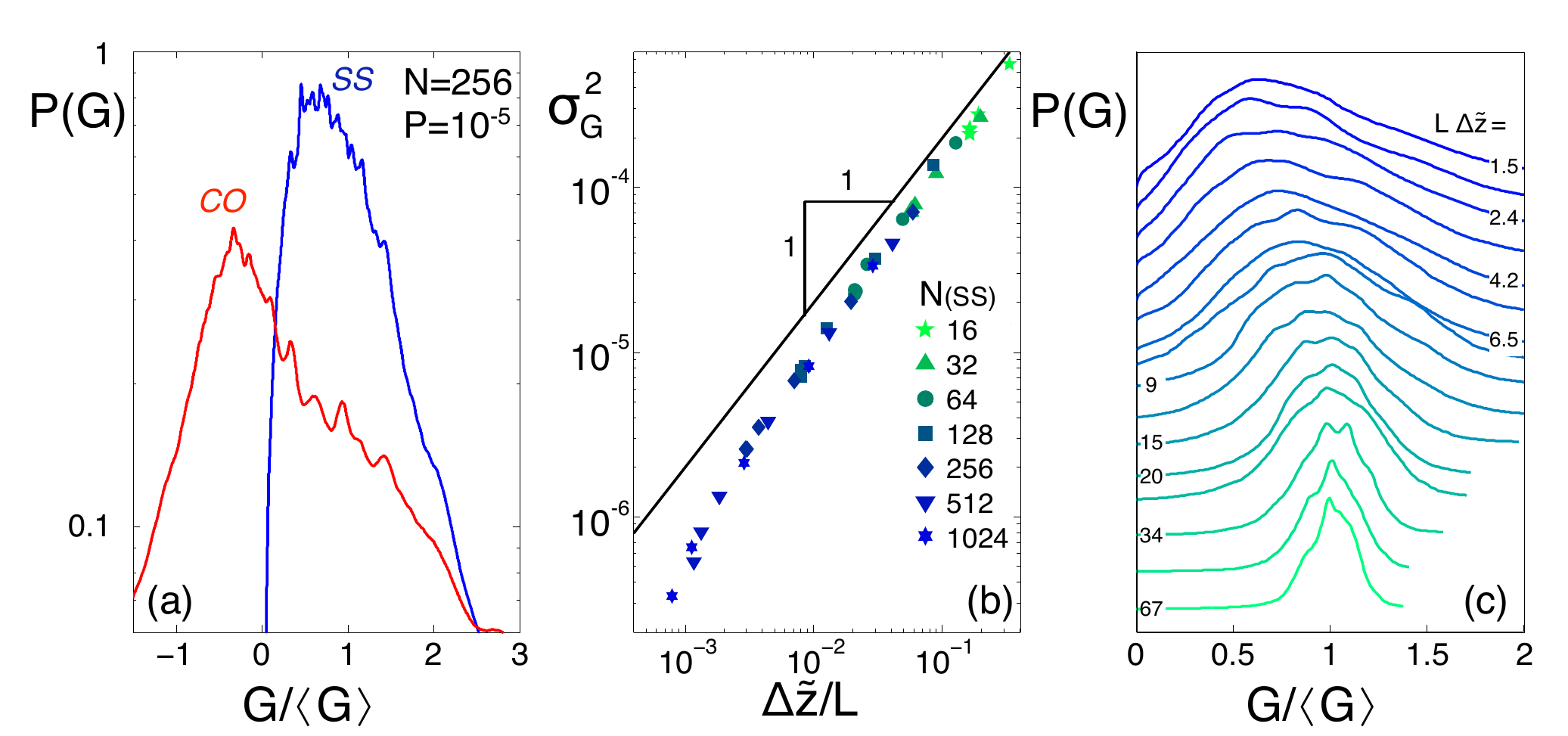}
\legende{Scaling}{(Color online) (a) The probability
distributions for $G$ of CO and SS packings differ qualitatively.
(b) Scaling of the variance $\langle (G-\langle G
\rangle)^2\rangle$ for SS packings reveals novel scaling. (c)
$P(G/\langle G \rangle)$ shows a systematic variation with $L
\Delta \tilde{z}$. }
\end{figure}

{\it Discussion ---} Improperly jammed CO packings dominate in the
critical, near jamming regime, whereas packings made by a shear
stabilized algorithm are strictly jammed: boundary conditions play
a crucial role in controlling the rigidity of packings, in
particular close to jamming.
%
In most experimental procedures, the creation history is richer
than homogeneously inflating particles, and involves the motion of
boundaries and shear
\cite{majmudar07,jenkins08,katgert10b,bi11,clusel09,cheng10,metayer11}
--- how does this relate to our scenario?
First, we note that in contrast to the 'shear jammed packing' of
Bi {\it et al.} \cite{bi11}, our CO and SS packings only exhibit
small contact anisotropies that vanish as $1/\sqrt{N}$
\cite{inprep}, and that CO packings exhibit similarly weak
anisotropies in the contact forces --- we use shear to stabilize,
rather than jam. Second, we note that the strong anisotropy of $G$
that we observe is reminiscent of fragility as introduced by Cates
{\it et al.}, although usually fragile states are defined as
having no resistance to shear in certain directions (i.e.,
$G\!=\!0$), while here we have $G\!<\!0$. Moreover, such fragility
typically arises due to the shear history of the system
\cite{cates98,bi11}. Nevertheless, it is conceivable that
protocols that do not explicitly perform shear stabilization
initially yield improperly jammed states, which then relax until
they reach a fragile state.

\begin{acknowledgements} We acknowledge discussions with
C. Goodrich, A. Liu, S. Nagel and Z. Zeravcic. SD-B acknowledges
funding from the Dutch physics foundation FOM and BPT from the
Netherlands Organization for Scientific Research.
\end{acknowledgements}

\bibliographystyle{apsrev}

\begin{thebibliography}{36}
\expandafter\ifx\csname natexlab\endcsname\relax\def\natexlab#1{#1}\fi
\expandafter\ifx\csname bibnamefont\endcsname\relax
  \def\bibnamefont#1{#1}\fi
\expandafter\ifx\csname bibfnamefont\endcsname\relax
  \def\bibfnamefont#1{#1}\fi
\expandafter\ifx\csname citenamefont\endcsname\relax
  \def\citenamefont#1{#1}\fi
\expandafter\ifx\csname url\endcsname\relax
  \def\url#1{\texttt{#1}}\fi
\expandafter\ifx\csname urlprefix\endcsname\relax\def\urlprefix{URL }\fi
\providecommand{\bibinfo}[2]{#2}
\providecommand{\eprint}[2][]{\url{#2}}

\bibitem[{\citenamefont{Majmudar et~al.}(2007)\citenamefont{Majmudar, Sperl,
  Luding, and Behringer}}]{majmudar07}
\bibinfo{author}{\bibfnamefont{T.~S.} \bibnamefont{Majmudar}},
  \bibinfo{author}{\bibfnamefont{M.}~\bibnamefont{Sperl}},
  \bibinfo{author}{\bibfnamefont{S.}~\bibnamefont{Luding}}, \bibnamefont{and}
  \bibinfo{author}{\bibfnamefont{R.~P.} \bibnamefont{Behringer}},
  \bibinfo{journal}{Phys.~Rev.~Lett.} \textbf{\bibinfo{volume}{98}},
  \bibinfo{pages}{058001} (\bibinfo{year}{2007}).

\bibitem[{\citenamefont{Lechenault et~al.}(2008)\citenamefont{Lechenault,
  Dauchot, Biroli, and Bouchaud}}]{lechenault08}
\bibinfo{author}{\bibfnamefont{F.}~\bibnamefont{Lechenault}},
  \bibinfo{author}{\bibfnamefont{O.}~\bibnamefont{Dauchot}},
  \bibinfo{author}{\bibfnamefont{G.}~\bibnamefont{Biroli}}, \bibnamefont{and}
  \bibinfo{author}{\bibfnamefont{J.~P.} \bibnamefont{Bouchaud}},
  \bibinfo{journal}{EPL} \textbf{\bibinfo{volume}{83}}, \bibinfo{pages}{46003}
  (\bibinfo{year}{2008}).

\bibitem[{\citenamefont{Jerkins et~al.}(2008)\citenamefont{Jerkins, Schr\"oter,
  Swinney, Senden, Saadatfar, and Aste}}]{jenkins08}
\bibinfo{author}{\bibfnamefont{M.}~\bibnamefont{Jerkins}},
  \bibinfo{author}{\bibfnamefont{M.}~\bibnamefont{Schr\"oter}},
  \bibinfo{author}{\bibfnamefont{H.~L.} \bibnamefont{Swinney}},
  \bibinfo{author}{\bibfnamefont{T.~J.} \bibnamefont{Senden}},
  \bibinfo{author}{\bibfnamefont{M.}~\bibnamefont{Saadatfar}},
  \bibnamefont{and} \bibinfo{author}{\bibfnamefont{T.}~\bibnamefont{Aste}},
  \bibinfo{journal}{Phys. Rev. Lett.} \textbf{\bibinfo{volume}{101}},
  \bibinfo{pages}{018301} (\bibinfo{year}{2008}).

\bibitem[{\citenamefont{Clusel et~al.}(2009)\citenamefont{Clusel, Corwin,
  Siemens, and Brujic}}]{clusel09}
\bibinfo{author}{\bibfnamefont{M.}~\bibnamefont{Clusel}},
  \bibinfo{author}{\bibfnamefont{E.~I.} \bibnamefont{Corwin}},
  \bibinfo{author}{\bibfnamefont{A.~O.~N.} \bibnamefont{Siemens}},
  \bibnamefont{and} \bibinfo{author}{\bibfnamefont{J.}~\bibnamefont{Brujic}},
  \bibinfo{journal}{Nature} \textbf{\bibinfo{volume}{460}},
  \bibinfo{pages}{611} (\bibinfo{year}{2009}).

\bibitem[{\citenamefont{Katgert and van Hecke}(2010)}]{katgert10b}
\bibinfo{author}{\bibfnamefont{G.}~\bibnamefont{Katgert}} \bibnamefont{and}
  \bibinfo{author}{\bibfnamefont{M.}~\bibnamefont{van Hecke}},
  \bibinfo{journal}{EPL} \textbf{\bibinfo{volume}{92}}, \bibinfo{pages}{34002}
  (\bibinfo{year}{2010}).

\bibitem[{\citenamefont{Cheng}(2010)}]{cheng10}
\bibinfo{author}{\bibfnamefont{X.}~\bibnamefont{Cheng}},
  \bibinfo{journal}{Phys. Rev. E} \textbf{\bibinfo{volume}{81}},
  \bibinfo{pages}{031301} (\bibinfo{year}{2010}).

\bibitem[{\citenamefont{M\'etayer et~al.}(2011)\citenamefont{M\'etayer, III,
  Radin, Swinney, and Schr\"oter}}]{metayer11}
\bibinfo{author}{\bibfnamefont{J.-F.} \bibnamefont{M\'etayer}},
  \bibinfo{author}{\bibfnamefont{D.~J.~S.} \bibnamefont{III}},
  \bibinfo{author}{\bibfnamefont{C.}~\bibnamefont{Radin}},
  \bibinfo{author}{\bibfnamefont{H.~L.} \bibnamefont{Swinney}},
  \bibnamefont{and}
  \bibinfo{author}{\bibfnamefont{M.}~\bibnamefont{Schr\"oter}},
  \bibinfo{journal}{EPL (Europhysics Letters)} \textbf{\bibinfo{volume}{93}},
  \bibinfo{pages}{64003} (\bibinfo{year}{2011}).

\bibitem[{\citenamefont{Bolton and Weaire}(1990)}]{bolton90}
\bibinfo{author}{\bibfnamefont{F.}~\bibnamefont{Bolton}} \bibnamefont{and}
  \bibinfo{author}{\bibfnamefont{D.}~\bibnamefont{Weaire}},
  \bibinfo{journal}{Phys. Rev. Lett.} \textbf{\bibinfo{volume}{65}},
  \bibinfo{pages}{3449} (\bibinfo{year}{1990}).

\bibitem[{\citenamefont{Durian}(1995)}]{durian95}
\bibinfo{author}{\bibfnamefont{D.~J.} \bibnamefont{Durian}},
  \bibinfo{journal}{Phys. Rev. Lett.} \textbf{\bibinfo{volume}{75}},
  \bibinfo{pages}{4780} (\bibinfo{year}{1995}).

\bibitem[{\citenamefont{O'Hern et~al.}(2003)\citenamefont{O'Hern, Silbert, Liu,
  and Nagel}}]{ohern03}
\bibinfo{author}{\bibfnamefont{C.~S.} \bibnamefont{O'Hern}},
  \bibinfo{author}{\bibfnamefont{L.~E.} \bibnamefont{Silbert}},
  \bibinfo{author}{\bibfnamefont{A.~J.} \bibnamefont{Liu}}, \bibnamefont{and}
  \bibinfo{author}{\bibfnamefont{S.~R.} \bibnamefont{Nagel}},
  \bibinfo{journal}{Phys.~Rev.~E} \textbf{\bibinfo{volume}{68}},
  \bibinfo{pages}{011306} (\bibinfo{year}{2003}).

\bibitem[{\citenamefont{van Hecke}(2010)}]{vanhecke10}
\bibinfo{author}{\bibfnamefont{M.}~\bibnamefont{van Hecke}},
  \bibinfo{journal}{J.~Phys.~Cond.~Matt.} \textbf{\bibinfo{volume}{22}},
  \bibinfo{pages}{033101} (\bibinfo{year}{2010}).

\bibitem[{\citenamefont{Liu and Nagel}(2010)}]{liu10a}
\bibinfo{author}{\bibfnamefont{A.~J.} \bibnamefont{Liu}} \bibnamefont{and}
  \bibinfo{author}{\bibfnamefont{S.~R.} \bibnamefont{Nagel}},
  \bibinfo{journal}{Annual Review of Condensed Matter Physics}
  \textbf{\bibinfo{volume}{1}}, \bibinfo{pages}{347} (\bibinfo{year}{2010}).

\bibitem[{\citenamefont{Zhang and Makse}(2005)}]{zhang05}
\bibinfo{author}{\bibfnamefont{H.~P.} \bibnamefont{Zhang}} \bibnamefont{and}
  \bibinfo{author}{\bibfnamefont{H.~A.} \bibnamefont{Makse}},
  \bibinfo{journal}{Phys. Rev. E} \textbf{\bibinfo{volume}{72}},
  \bibinfo{pages}{011301} (\bibinfo{year}{2005}).

\bibitem[{\citenamefont{Silbert et~al.}(2005)\citenamefont{Silbert, Liu, and
  Nagel}}]{silbert05}
\bibinfo{author}{\bibfnamefont{L.~E.} \bibnamefont{Silbert}},
  \bibinfo{author}{\bibfnamefont{A.~J.} \bibnamefont{Liu}}, \bibnamefont{and}
  \bibinfo{author}{\bibfnamefont{S.~R.} \bibnamefont{Nagel}},
  \bibinfo{journal}{Phys.~Rev.~Lett.} \textbf{\bibinfo{volume}{95}},
  \bibinfo{pages}{098301} (\bibinfo{year}{2005}).


\bibitem[{\citenamefont{Ellenbroek et~al.}(2006)\citenamefont{Ellenbroek,
  Somfai, van Hecke, and van Saarloos}}]{ellenbroek06}
\bibinfo{author}{\bibfnamefont{W.~G.} \bibnamefont{Ellenbroek}},
  \bibinfo{author}{\bibfnamefont{E.}~\bibnamefont{Somfai}},
  \bibinfo{author}{\bibfnamefont{M.}~\bibnamefont{van Hecke}},
  \bibnamefont{and} \bibinfo{author}{\bibfnamefont{W.}~\bibnamefont{van
  Saarloos}}, \bibinfo{journal}{Phys.~Rev.~Lett.}
  \textbf{\bibinfo{volume}{97}}, \bibinfo{pages}{258001}
  (\bibinfo{year}{2006}).

\bibitem[{\citenamefont{Tighe and Vlugt}(2011)}]{tighe11b}
\bibinfo{author}{\bibfnamefont{B.~P.} \bibnamefont{Tighe}} \bibnamefont{and}
  \bibinfo{author}{\bibfnamefont{T.~J.~H.} \bibnamefont{Vlugt}},
  \bibinfo{journal}{Journal of Statistical Mechanics: Theory and Experiment} p.
  \bibinfo{pages}{P04002} (\bibinfo{year}{2011}).

\bibitem[{\citenamefont{Tighe}(2011)}]{tighe11}
\bibinfo{author}{\bibfnamefont{B.~P.} \bibnamefont{Tighe}},
  \bibinfo{journal}{Phys. Rev. Lett.} \textbf{\bibinfo{volume}{107}},
  \bibinfo{pages}{158303} (\bibinfo{year}{2011}).



\bibitem{footnote1} The distinction between
CO and SS packings is comparable to the difference between what
Torquato {\it et al.} refer to as collectively and strictly jammed
packings, although these concepts are defined for hard particles
\cite{torquato01}.


\bibitem[{\citenamefont{Torquato and Stillinger}(2001)}]{torquato01}
\bibinfo{author}{\bibfnamefont{S.}~\bibnamefont{Torquato}} \bibnamefont{and}
  \bibinfo{author}{\bibfnamefont{F.}~\bibnamefont{Stillinger}},
  \bibinfo{journal}{J. Phys. Chem B} \textbf{\bibinfo{volume}{105}},
  \bibinfo{pages}{11849} (\bibinfo{year}{2001}).

\bibitem[{\citenamefont{O'Hern et~al.}(2002)\citenamefont{O'Hern, Langer, Liu,
  and Nagel}}]{ohern02}
\bibinfo{author}{\bibfnamefont{C.~S.} \bibnamefont{O'Hern}},
  \bibinfo{author}{\bibfnamefont{S.~A.} \bibnamefont{Langer}},
  \bibinfo{author}{\bibfnamefont{A.~J.} \bibnamefont{Liu}}, \bibnamefont{and}
  \bibinfo{author}{\bibfnamefont{S.~R.} \bibnamefont{Nagel}},
  \bibinfo{journal}{Phys.~Rev.~Lett.} \textbf{\bibinfo{volume}{88}},
  \bibinfo{pages}{075507} (\bibinfo{year}{2002}).

\bibitem[{\citenamefont{Maloney and Lema\^\i{}tre}(2006)}]{maloney06}
\bibinfo{author}{\bibfnamefont{C.~E.} \bibnamefont{Maloney}} \bibnamefont{and}
  \bibinfo{author}{\bibfnamefont{A.}~\bibnamefont{Lema\^\i{}tre}},
  \bibinfo{journal}{Phys. Rev. E} \textbf{\bibinfo{volume}{74}},
  \bibinfo{pages}{016118} (\bibinfo{year}{2006}).

\bibitem[{\citenamefont{Somfai et~al.}(2007)\citenamefont{Somfai, van Hecke,
  Ellenbroek, Shundyak, and van Saarloos}}]{somfai07}
\bibinfo{author}{\bibfnamefont{E.}~\bibnamefont{Somfai}},
  \bibinfo{author}{\bibfnamefont{M.}~\bibnamefont{van Hecke}},
  \bibinfo{author}{\bibfnamefont{W.~G.} \bibnamefont{Ellenbroek}},
  \bibinfo{author}{\bibfnamefont{K.}~\bibnamefont{Shundyak}}, \bibnamefont{and}
  \bibinfo{author}{\bibfnamefont{W.}~\bibnamefont{van Saarloos}},
  \bibinfo{journal}{Phys.~Rev.~E} \textbf{\bibinfo{volume}{75}},
  \bibinfo{pages}{020301(R)} (\bibinfo{year}{2007}).

\bibitem[{\citenamefont{Wyart et~al.}(2008)\citenamefont{Wyart, Liang, Kabla,
  and Mahadevan}}]{wyart08}
\bibinfo{author}{\bibfnamefont{M.}~\bibnamefont{Wyart}},
  \bibinfo{author}{\bibfnamefont{H.}~\bibnamefont{Liang}},
  \bibinfo{author}{\bibfnamefont{A.}~\bibnamefont{Kabla}}, \bibnamefont{and}
  \bibinfo{author}{\bibfnamefont{L.}~\bibnamefont{Mahadevan}},
  \bibinfo{journal}{Phys. Rev. Lett.} \textbf{\bibinfo{volume}{101}},
  \bibinfo{pages}{215501} (\bibinfo{year}{2008}).

\bibitem[{\citenamefont{Zaccone and Scossa-Romano}(2011)}]{zaccone11}
\bibinfo{author}{\bibfnamefont{A.}~\bibnamefont{Zaccone}} \bibnamefont{and}
  \bibinfo{author}{\bibfnamefont{E.}~\bibnamefont{Scossa-Romano}},
  \bibinfo{journal}{Phys. Rev. B} \textbf{\bibinfo{volume}{83}},
  \bibinfo{pages}{184205} (\bibinfo{year}{2011}).

\bibitem{privcoms}W.~G.~Ellenbroek, C.~S.~O'Hern (private communication).

\bibitem[{\citenamefont{Schreck and O'Hern}(2010)}]{schreck10}
\bibinfo{author}{\bibfnamefont{C.}~\bibnamefont{Schreck}} \bibnamefont{and}
  \bibinfo{author}{\bibfnamefont{C.}~\bibnamefont{O'Hern}},
  {\it {\bibinfo{title}{Computational methods to study jammed systems}}}
  (\bibinfo{publisher}{Cambridge University Press}, \bibinfo{year}{2010}).

\bibitem{notemin} To obtain positive $G$, in principle one only
requires the sign of the curvature of $U$, $\partial^2 U /\partial
\gamma^2$, to be positive. At a minumum of $U$, $\partial U
/\partial \gamma = 0$ as well, leading to states with zero
residual shear stress.

\bibitem[{\citenamefont{Bitzek et~al.}(2006)\citenamefont{Bitzek, Koskinen,
  G\"ahler, Moseler, and Gumbsch}}]{fire}
\bibinfo{author}{\bibfnamefont{E.}~\bibnamefont{Bitzek}},
  \bibinfo{author}{\bibfnamefont{P.}~\bibnamefont{Koskinen}},
  \bibinfo{author}{\bibfnamefont{F.}~\bibnamefont{G\"ahler}},
  \bibinfo{author}{\bibfnamefont{M.}~\bibnamefont{Moseler}}, \bibnamefont{and}
  \bibinfo{author}{\bibfnamefont{P.}~\bibnamefont{Gumbsch}},
  \bibinfo{journal}{Phys. Rev. Lett.} \textbf{\bibinfo{volume}{97}},
  \bibinfo{pages}{170201} (\bibinfo{year}{2006}).


\bibitem{footnote3b} We stress here that these anisotropies pertain to
individual packings --- ensembles of CO or SS packings are
isotropic.

\bibitem[{\citenamefont{O'Hern et~al.}(2004)\citenamefont{O'Hern, Silbert, Liu,
  and Nagel}}]{ohern04}
\bibinfo{author}{\bibfnamefont{C.~S.} \bibnamefont{O'Hern}},
  \bibinfo{author}{\bibfnamefont{L.~E.} \bibnamefont{Silbert}},
  \bibinfo{author}{\bibfnamefont{A.~J.} \bibnamefont{Liu}}, \bibnamefont{and}
  \bibinfo{author}{\bibfnamefont{S.~R.} \bibnamefont{Nagel}},
  \bibinfo{journal}{Phys. Rev. E} \textbf{\bibinfo{volume}{70}},
  \bibinfo{pages}{043302} (\bibinfo{year}{2004}).


\bibitem{Pnote}For harmonic particles, $\Delta z \sim P^{1/2}$
\cite{vanhecke10,liu10a}


\bibitem[{\citenamefont{Tkachenko and Witten}(1999)}]{tkachenko99}
\bibinfo{author}{\bibfnamefont{A.~V.} \bibnamefont{Tkachenko}}
  \bibnamefont{and} \bibinfo{author}{\bibfnamefont{T.~A.}
  \bibnamefont{Witten}}, \bibinfo{journal}{Phys.~Rev.~E}
  \textbf{\bibinfo{volume}{60}}, \bibinfo{pages}{687} (\bibinfo{year}{1999}).

\bibitem[{\citenamefont{Wyart et~al.}(2005)\citenamefont{Wyart, Nagel, and
  Witten}}]{wyart05}
\bibinfo{author}{\bibfnamefont{M.}~\bibnamefont{Wyart}},
  \bibinfo{author}{\bibfnamefont{S.~R.} \bibnamefont{Nagel}}, \bibnamefont{and}
  \bibinfo{author}{\bibfnamefont{T.~A.} \bibnamefont{Witten}},
  \bibinfo{journal}{EPL} \textbf{\bibinfo{volume}{72}},
  \bibinfo{pages}{486} (\bibinfo{year}{2005}).

\bibitem{inprep} S. Dagois-Bohy, B. P. Tighe and M. van Hecke, in
preparation



\bibitem[{\citenamefont{Alexander}(1998)}]{alexander}
\bibinfo{author}{\bibfnamefont{S.}~\bibnamefont{Alexander}},
  \bibinfo{journal}{Phys.~Rep} \textbf{\bibinfo{volume}{296}},
  \bibinfo{pages}{65} (\bibinfo{year}{1998}).



\bibitem{note_rattler} Here $N$ denotes the number of particles
after a small fraction of non-force bearing particles, or
``rattlers,'' have been removed.

\bibitem{CSL} C. Goodrich, S. R. Nagel and A. J. Liu,
arXiv:1204.4915.


\bibitem[{\citenamefont{Henkes et~al.}(2007)\citenamefont{Henkes, O'Hern, and
  Chakraborty}}]{henkes07}
\bibinfo{author}{\bibfnamefont{S.}~\bibnamefont{Henkes}},
  \bibinfo{author}{\bibfnamefont{C.~S.} \bibnamefont{O'Hern}},
  \bibnamefont{and}
  \bibinfo{author}{\bibfnamefont{B.}~\bibnamefont{Chakraborty}},
  \bibinfo{journal}{Phys.~Rev.~Lett.} \textbf{\bibinfo{volume}{99}},
  \bibinfo{pages}{038002} (\bibinfo{year}{2007}).

\bibitem[{\citenamefont{Tighe et~al.}(2008)\citenamefont{Tighe, van Eerd, and
  Vlugt}}]{tighe08b}
\bibinfo{author}{\bibfnamefont{B.~P.} \bibnamefont{Tighe}},
  \bibinfo{author}{\bibfnamefont{A.~R.~T.} \bibnamefont{van Eerd}},
  \bibnamefont{and} \bibinfo{author}{\bibfnamefont{T.~J.~H.}
  \bibnamefont{Vlugt}}, \bibinfo{journal}{Phys.~Rev.~Lett.}
  \textbf{\bibinfo{volume}{100}}, \bibinfo{pages}{238001}
  (\bibinfo{year}{2008}).

\bibitem[{\citenamefont{Ellenbroek et~al.}(2009)\citenamefont{Ellenbroek, van
  Hecke, and van Saarloos}}]{ellenbroek09}
\bibinfo{author}{\bibfnamefont{W.~G.} \bibnamefont{Ellenbroek}},
  \bibinfo{author}{\bibfnamefont{M.}~\bibnamefont{van Hecke}},
  \bibnamefont{and} \bibinfo{author}{\bibfnamefont{W.}~\bibnamefont{van
  Saarloos}}, \bibinfo{journal}{Phys. Rev. E} \textbf{\bibinfo{volume}{80}},
  \bibinfo{pages}{061307} (\bibinfo{year}{2009}).

\bibitem[{\citenamefont{Heussinger et~al.}(2010)\citenamefont{Heussinger,
  Chaudhuri, and Barrat}}]{heussinger10}
\bibinfo{author}{\bibfnamefont{C.}~\bibnamefont{Heussinger}},
  \bibinfo{author}{\bibfnamefont{P.}~\bibnamefont{Chaudhuri}},
  \bibnamefont{and} \bibinfo{author}{\bibfnamefont{J.-L.}
  \bibnamefont{Barrat}}, \bibinfo{journal}{Soft Matter}
  \textbf{\bibinfo{volume}{6}} (\bibinfo{year}{2010}).

\bibitem[{\citenamefont{Tighe et~al.}(2010)\citenamefont{Tighe, Woldhuis,
  Remmers, van Saarloos, and van Hecke}}]{tighe10c}
\bibinfo{author}{\bibfnamefont{B.~P.} \bibnamefont{Tighe}},
  \bibinfo{author}{\bibfnamefont{E.}~\bibnamefont{Woldhuis}},
  \bibinfo{author}{\bibfnamefont{J.~J.~C.} \bibnamefont{Remmers}},
  \bibinfo{author}{\bibfnamefont{W.}~\bibnamefont{van Saarloos}},
  \bibnamefont{and} \bibinfo{author}{\bibfnamefont{M.}~\bibnamefont{van
  Hecke}}, \bibinfo{journal}{Phys. Rev. Lett.} \textbf{\bibinfo{volume}{105}},
  \bibinfo{pages}{088303} (\bibinfo{year}{2010}).

\bibitem{carlpriv} C. Goodrich, A. J. Liu, S. R. Nagel, Priv.
Comm.

\bibitem[{\citenamefont{Bi et~al.}(2011)\citenamefont{Bi, Zhang, Chakraborty,
  and Behringer}}]{bi11}
\bibinfo{author}{\bibfnamefont{D.}~\bibnamefont{Bi}},
  \bibinfo{author}{\bibfnamefont{J.}~\bibnamefont{Zhang}},
  \bibinfo{author}{\bibfnamefont{B.}~\bibnamefont{Chakraborty}},
  \bibnamefont{and} \bibinfo{author}{\bibfnamefont{R.~P.}
  \bibnamefont{Behringer}}, \bibinfo{journal}{Nature}
  \textbf{\bibinfo{volume}{480}}, \bibinfo{pages}{355} (\bibinfo{year}{2011}).

\bibitem[{\citenamefont{Cates et~al.}(1998)\citenamefont{Cates, Wittmer,
  Bouchaud, and Claudin}}]{cates98}
\bibinfo{author}{\bibfnamefont{M.~E.} \bibnamefont{Cates}},
  \bibinfo{author}{\bibfnamefont{J.~P.} \bibnamefont{Wittmer}},
  \bibinfo{author}{\bibfnamefont{J.-P.} \bibnamefont{Bouchaud}},
  \bibnamefont{and} \bibinfo{author}{\bibfnamefont{P.}~\bibnamefont{Claudin}},
  \bibinfo{journal}{Phys. Rev. Lett.} \textbf{\bibinfo{volume}{81}},
  \bibinfo{pages}{1841} (\bibinfo{year}{1998}).

\end{thebibliography}

~\newline

~\newline

{\large Supp Material}

{\it Extended Hessian ---} A packing's linear response to shear
can be expressed as a matrix equation. Let us introduce $|u\rangle
= |\lbrace u^x_i, u^y_i \rbrace_{i=1}^{N} \rangle$, $|q \rangle =
|\lbrace u^x_i, u^y_i \rbrace_{i=1}^{N} , \gamma,\theta \rangle$
and
\begin{equation}
K_{mn}^0 = \frac{\partial^2 U}{\partial u_m \, \partial u_n}
\,\,\,\,\, {\rm and} \,\,\,\, K_{mn}= \frac{\partial^2 U}{\partial
q_m \, \partial q_n}
 \,,
\end{equation}
evaluated at the coordinates corresponding to a packing. $\hat
K^0$ is the usual Hessian or stiffness matrix, while the
``extended Hessian'' $\hat K$, introduced in Ref.~\cite{tighe11},
includes the dependence on $\gamma$ and $\theta$. The response to
imposed strain is the solution to ${\hat K}^0 | u\rangle =
|F_\Gamma \rangle $, where $|F_\Gamma \rangle$ is an apparent
force felt by particles involved in boundary-crossing contacts
when the lattice vectors are distorted. It comprises the first
$2N$ components of  ${\hat K} |\Gamma \rangle $, where $|\Gamma
\rangle = |\lbrace 0 \rbrace_{2N}, \gamma,\theta \rangle$. The
quadratic term in the change in potential energy, which governs
linear stability, is then $\Delta U/V = (1/2V) \langle q | \hat K | q
\rangle \equiv (1/2) G(\theta) \gamma^2$, where $V$ is the volume,
so that
\begin{equation}
G(\theta) =  \langle q | \hat K | q \rangle /
\gamma^2 V~. \label{Gexp}
\end{equation}

From Eq.~(\ref{Gexp}) it immediately follows that $G(\theta)$ is
positive definite for all $\theta$  if all eigenvalues of $\hat K$
are {\it positive} (excluding the trivial zero energy
translational modes).

We finally note that Eq.~(\ref{Gexp}) requires the extended
Hessian $\hat K$, which includes shear degrees of freedom
\cite{tighe11}. To relate $G$ to the usual Hessian $\hat K^0$, we
follow Ref.~\cite{maloney06} and write $G\gamma^2/2V = W_{\rm a} -
W_{\rm na}$, where $W_{\rm a} > 0$ is the work done in affinely
displacing the particles and the box. The actual particle
displacements $|u\rangle$ have a non-affine contribution $|u_{\rm
na}\rangle = |u \rangle - |u_{\rm a}\rangle$ that reduces the
deformation energy by $W_{\rm na} = \langle u_{\rm na} | K^0 |
u_{\rm na} \rangle / 2$. $W_{\rm na}> 0$ if the spectrum of $K^0$
is non-negative, but $G$ can still be negative if $W_{\rm na} >
W_{\rm a}$. --- hence there is no simple relation between the sign
of $G$ and the eigenvalues of the usual Hessian $\hat K^0$.

{\it Contact Numbers for  $P\rightarrow 0$ ---} Based on our
counting, the isostatic number of contacts, $C_{\rm iso}$, equals
$2N-2$ for the CO ensemble, and precisely $2N$ for the SS
ensemble. In Fig.~7 we show our results for the excess contact
numbers $C^+$, defined as the difference between the actual number
of contacts $C$ and the respective isostatic  value $C_{\rm iso}$.
For both the CO and SS ensembles, the number of excess contacts
reaches one in the limit of vanishing pressure: so for 2D CO
ensembles, the number of contacts reaches $2N-1$, and for 2D SS
ensembles, the number of contacts reaches $2N+1$.

\begin{figure}[b]
\includegraphics[clip,width=0.9\linewidth]{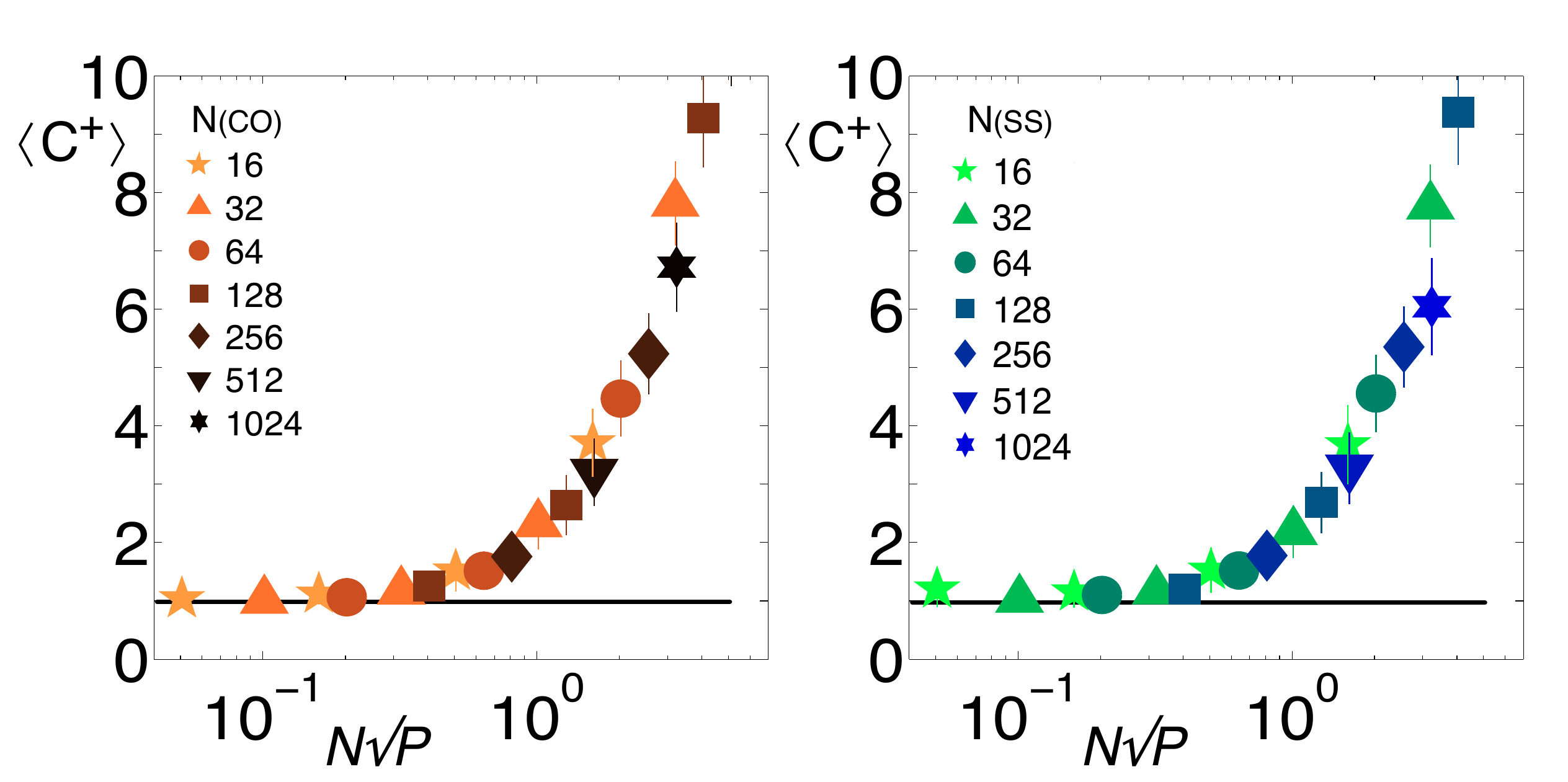}
\centering \legende{G}{(Color online) Scaling of the excess
contact number of contacts, $C+ = C-C_{\rm iso}$ for (a) the CO
and (b) the SS ensemble.}
\end{figure}

\end{document}